\documentclass[12pt,preprint]{aastex}

\shorttitle{Temperature Anisotropy in a Shocked Plasma}
\shortauthors{Liu et al.}

\begin{document}

\title{Temperature Anisotropy in a Shocked Plasma: Mirror-Mode
Instabilities in the Heliosheath}

\author{Y. Liu\altaffilmark{1,2}, J. D. Richardson\altaffilmark{1,2},
J. W. Belcher\altaffilmark{1}, and J. C. Kasper\altaffilmark{1}}

\altaffiltext{1}{Kavli Institute for Astrophysics and Space Research,
Massachusetts Institute of Technology, Cambridge, MA 02139, USA;
liuxying@mit.edu.}

\altaffiltext{2}{State Key Laboratory of Space Weather, Chinese
Academy of Sciences, Beijing 100080, China.}

\begin{abstract}
We show that temperature anisotropies induced at a shock can account
for interplanetary and planetary bow shock observations. Shocked
plasma with enhanced plasma beta is preferentially unstable to the
mirror mode instability downstream of a quasi-perpendicular shock and
to the firehose instability downstream of a quasi-parallel shock,
consistent with magnetic fluctuations observed downstream of a large
variety of shocks. Our theoretical analysis of the solar wind
termination shock suggests that the magnetic holes observed by
Voyager 1 in the heliosheath are produced by the mirror mode
instability. The results are also of astrophysical interest,
providing an energy source for plasma heating.
\end{abstract}


\section{Introduction}
Planetary bow shocks and interplanetary shocks serve as a unique
laboratory for the study of shock waves in collisionless plasmas.
Observations of these shocks usually show that ion distributions are
anisotropic with respect to the background magnetic field downstream
of the shocks. Mirror mode waves associated with this anisotropy are
observed downstream of quasi-perpendicular shocks (defined by the
angle between the shock normal and the upstream magnetic field
$\theta_{Bn}> 45^{\circ}$) when the plasma beta is high ($\beta>1$)
\citep{kaufmann70, tsurutani92, violante95, bavassano98, liu06a}.
Mirror mode waves do not grow in low beta regions where the ion
cyclotron mode dominates \citep[e.g.,][]{anderson94, czaykowska01}.
Magnetic fluctuations downstream of quasi-parallel shocks
($\theta_{Bn}< 45^{\circ}$) have not been identified in detail, but
hybrid simulations show that the firehose instability can occur
downstream of these shocks for certain ranges of upstream Alfv\'{e}n
Mach number ($M_{\rm A}$) and plasma beta, for instance, $M_{\rm
A}\geq 3$ at $\beta\sim 0.5$ \citep{kan83, krauss91}; observations
seem to confirm this point \citep[e.g.,][]{greenstadt79,
bavassano00}. Quasi-perpendicular shocks are characterized by a sharp
increase in the magnetic field strength, but quasi-parallel shocks
are often more turbulent, with shock ramps containing large-amplitude
waves which spread upstream and downstream.

The recent crossing of the termination shock (TS) by Voyager 1 (V1)
\citep{burlaga05, decker05, gurnett05, stone05} provides an
opportunity to study shocks and shock-induced waves in the
heliosheath. The sharp increase in the magnetic field strength across
the TS and the downstream magnetic field configuration suggest that
the TS is quasi-perpendicular \citep{burlaga05}. As in planetary
magnetosheaths downstream of a quasi-perpendicular bow shock, the
heliosheath shows compressive magnetic fluctuations in the form of
magnetic holes \citep{burlaga06a, burlaga06b}.

We propose a theoretical explanation for the temperature anisotropies
and associated instabilities induced at a shock. Based on the
theoretical analysis, we show that the magnetic holes observed in the
heliosheath could be mirror mode fluctuations. The present results
also provide a prototype for understanding shocks in various
astrophysical contexts, such as gamma-ray bursts, supernova
explosions and active galactic nuclei.

\section{Theory}
An anisotropic ion distribution requires the use of a pressure tensor
in the magnetohydrodynamic (MHD) equations. Observations show that
the shock structure and dynamics depend on the shock geometry and
upstream $M_{\rm A1}$ and $\beta_1$, so we examine the temperature
anisotropy $A_2=T_{\perp 2}/T_{\parallel 2}$ as a function of these
parameters, where $T_{\perp}$ and $T_{\parallel}$ are the plasma
temperature perpendicular and parallel to the magnetic field. The
subscripts 1 and 2 indicate physical parameters upstream and
downstream of a shock. Assuming a bi-Maxwellian plasma, we write the
jump conditions across a shock as \citep{hudson70}
\begin{equation}
\left[ B_n \right] = 0,
\end{equation}
\begin{equation}
\left[ \rho v_n \right] = 0,
\end{equation}
\begin{equation}
\left[ v_n \textbf{B}_t - \textbf{v}_t B_n \right] = 0,
\end{equation}
\begin{equation}
\left[ P_{\perp} + (P_{\parallel}-P_{\perp})\frac{B_n^2}{B^2} +
\frac{B_t^2 - B_n^2}{2\mu_0} + \rho v_n^2 \right] = 0,
\end{equation}
\begin{equation}
\left[\frac{B_n\textbf{B}_t}{\mu_0}\left(\frac{P_{\parallel}-
P_{\perp}}{B^2/\mu_0}-1\right) + \rho v_n\textbf{v}_t \right] = 0,
\end{equation}
\begin{equation}
\left[\rho v_n\left(\frac{2P_{\perp}}{\rho} +
\frac{P_{\parallel}}{2\rho} + \frac{v^2}{2} + \frac{B^2_t}{\mu_0\rho}
\right) + \frac{B_n^2v_n}{B^2}(P_{\parallel} - P_{\perp}) -
\frac{(\textbf{B}_t\cdot\textbf{v}_t)B_n}{\mu_0}
\left(1-\frac{P_{\parallel}-P_{\perp}}{B^2/\mu_0} \right) \right]=0,
\end{equation}
where $\mu_0$ is the permeability of vacuum, $\rho$ is the plasma
density, and \textbf{v} and \textbf{B} are the plasma velocity and
magnetic field with subscripts $t$ and $n$ denoting the tangential
and normal components with respect to the shock surface. The velocity
is measured in the shock frame. The square brackets indicate the
difference between the preshock (1) and postshock (2) states. The
perpendicular and parallel plasma pressures are defined as $P_{\perp}
= \rho k_{\rm B}T_{\perp}/m_p$ and $P_{\parallel} = \rho k_{\rm
B}T_{\parallel}/m_p$, where $k_{\rm B}$ and $m_p$ represent the
Boltzmann constant and proton mass, respectively. For simplicity, we
assume that the bulk velocity is parallel to the shock normal. The
components of the magnetic field upstream of the shock are given by
$B_{n1} = B_1 \cos\theta_{Bn}$ and $B_{t1} = B_1 \sin\theta_{Bn}$.
Different forms of the solutions to these equations have been derived
based on various assumptions \citep{chao95, erkaev00}. The focus of
the present analysis is to investigate under what conditions the
shocked plasma is unstable to the thermal anisotropy instabilities.

The thermal anisotropy serves a free energy source which can feed
magnetic fluctuations when it exceeds certain threshold conditions.
The thresholds can be expressed as $A = 1 - 2/\beta_{\parallel}$ for
the firehose instability \citep{parker58}, the lower bound of the
temperature anisotropy, and $A = 1 + 1/\beta_{\perp}$ for the mirror
mode instability \citep{chan58, hasegawa69}, the upper bound. The ion
cyclotron instability, competing with the mirror mode, has the onset
condition $A = 1+ 0.64/\beta_{\parallel}^{0.41}$ \citep{gary97}. Here
the perpendicular and parallel plasma betas are defined as
$\beta_{\perp} = P_{\perp}/(B^2/2\mu_0)$ and $\beta_{\parallel} =
P_{\parallel}/(B^2/2\mu_0)$.

For a perpendicular shock, equations~(2) and (3) give the shock
strength
\begin{equation}
r_s = \frac{\rho_2}{\rho_1} = \frac{v_1}{v_2} = \frac{B_2}{B_1}.
\end{equation}
The temperature anisotropy $A_2$ can be obtained from equations~(4),
(5) and (6),
\begin{equation}
A_2 = \frac{\frac{3A_1\beta_1}{2A_1 + 1} + 2M_{A1}^2(1-1/r_s) + 1 -
r_s^2}{\frac{3A_1 \beta_1}{2A_1 + 1}(4r_s - 4 + r_s/A_1) +
2M_{A1}^2(r_s + 3/r_s -4) + 4(r_s -1)},
\end{equation}
where $\beta = (2P_{\perp} + P_{\parallel})/(3B^2/2\mu_0)$. To
compare with the thresholds, we derive $\beta_{\perp 2}$ from
equation~(4) as
\begin{equation}
\beta_{\perp 2} = \frac{3A_1\beta_1}{(2A_1 + 1)r_s^2} +
\frac{2M_{A1}^2}{r_s^2}(1-1/r_s) - 1 + \frac{1}{r_s^2}.
\end{equation}
For $2M_{A1}^2 \gg \beta_1$, we have $\beta_{\perp 2} \sim
2M_{A1}^2/r_s^2$, so the shocked plasma would have a high beta
independent of $\beta_1$, consistent with the findings in the
magnetosheaths of outer planets \citep{russell90}. As can be seen
from the mirror-mode threshold, high values of $\beta_{\perp 2}$
favor the onset of the mirror mode instability; similarly, $A_2 \sim
1/(r_s-3)$ under the same condition, so $A_2 \geq 1$ since the shock
strength cannot be larger than 4, also favoring the mirror mode
onset.

Figure~1 shows the temperature anisotropy $A_2$ for $r_s=3$ and
$A_1=1$ as a function of $M_{A1}$ and $\beta_1$. Only values of
$0\leq A_2 \leq 2$ are shown. The entropy, $S = \frac{1}{2}k_{\rm
B}\ln(P_{\perp}^2P_{\parallel}/\rho^5)$, is required to increase
across a shock by the second law of thermodynamics; the region which
breaks the entropy principle is show as ``Forbidden'' in Figure~1.
The majority of the allowed area has $A_2 \geq 1$ as expected and is
preferentially unstable to the mirror mode instability. The mirror
mode has a lower threshold than the cyclotron mode in this plasma
regime, so the temperature anisotropy would be quickly reduced by the
mirror mode before the ion cyclotron mode could develop. The TS with
$\beta_1\simeq 32.8$ and $M_{A1}\simeq 16.3$ (see \S 3), indicated by
the plus sign, is located slightly above the mirror mode threshold.
Interplanetary and planetary bow shocks have a large variation in
$M_{A1}$ and $\beta_1$; many of them would also be above the mirror
mode threshold as indicated by the large area with $A_2 \geq 1$.
Consequently, mirror mode instabilities should be a frequent feature
downstream of quasi-perpendicular shocks.

For a parallel shock, the magnetic field does not go into the
expression of the shock strength. The temperature anisotropy
\begin{equation}
A_2 = \frac{\frac{3\beta_1 r_s}{2A_1 + 1}(A_1 + \frac{3}{2} -
\frac{3}{2r_s}) + M_{A1}^2(r_s + 2/r_s -3)}{\frac{3\beta_1}{2A_1 + 1}
+ 2M_{A1}^2(1-1/r_s)},
\end{equation}
and the downstream parallel plasma beta
\begin{equation}
\beta_{\parallel 2} = \frac{3\beta_1}{2A_1+1} + 2M_{A1}^2(1-1/r_s).
\end{equation}
For $2M_{A1}^2 \gg \beta_1$, $\beta_{\parallel 2} \sim 2M_{A1}^2$,
which makes the thresholds close to 1, and $A_2 \sim (r_s - 2)/2 \leq
1$, giving favorable conditions for the onset of the firehose
instability. Figure~2 displays the temperature anisotropy $A_2$ over
various upstream conditions for $r_s =3$ and $A_1=1$. Compared with
Figure~1, higher values of $M_{A1}$ at a given $\beta_1$ would
otherwise lead to smaller $A_2$. Most of the area shown has $A_2\leq
1$ as expected for a quasi-parallel shock and is unstable to the
firehose instability. The TS would induce firehose instabilities in
the downstream plasma if it were quasi-parallel, as indicated by its
location in the plot. Many quasi-parallel interplanetary and
planetary bow shocks will also give rise to firehose instabilities as
implied by the large area with $A_2\leq 1$.

Observations show that a quasi-parallel bow shock becomes unsteady as
the Alfv\'{e}n Mach number $M_{\rm A1}$ exceeds $\sim 3$ for $\beta_1
\sim 0.5$ \citep{greenstadt79} and is often associated with large
transverse magnetic fluctuations \citep{bavassano00, czaykowska01}. A
closer look at Figure~2 gives $A_2\simeq 0.97$ at $M_{A1} = 2$ for
$\beta_1=0.5$, a noticeable thermal anisotropy but not large enough
to exceed the firehose onset; at $M_{A1} = 3$, the thermal anisotropy
rises above the firehose threshold, generating firehose instabilities
which will disturb the shock structure. The firehose instability
arises from the fast MHD mode and produces large-amplitude transverse
waves. The results from this simple calculation agree with hybrid
simulations \citep{kan83, krauss91}.

\section{Magnetic Fluctuations in the Heliosheath}
V1 crossed the TS on 2004 December 16 (day 351) at 94 AU and is
making the first measurements in the heliosheath. The magnetic
fluctuations in the heliosheath are characterized by a series of
depressions in the field magnitude which have been called magnetic
holes \citep{burlaga06a, burlaga06b}. These fluctuations are similar
to those observed downstream of quasi-perpendicular interplanetary
and planetary bow shocks that have been identified as mirror mode
structures \citep{kaufmann70, tsurutani92, violante95, bavassano98}.

The TS is expected to be highly oblique with $\theta_{Bn} \sim
86^{\circ}$ at V1, so we use equations~(8) and (9) to determine
whether mirror mode instabilities occur in the heliosheath. MHD
simulations give the average preshock plasma density $n_1\simeq
8\times 10^{-4}$ cm$^{-3}$, speed $v_1\simeq 380$ km s$^{-1}$, and
temperature $T_1\simeq 5.4\times 10^5$ K \citep{whang04}, which
yields $M_{A1}\simeq 16.3$, $\beta_1\simeq 32.8$ combined with the
observed field strength $B_1\simeq 0.03$ nT. The shock strength $r_s$
is $\sim 3$ estimated from the jump in the field magnitude
\citep{burlaga05} and from the spectral slope of TS accelerated
particles \citep{stone05}. The typical thermal anisotropy of the
solar wind is $A \sim 0.7$ near the Earth \citep{liu06b}. Expansion
of the solar wind would further decrease the value of $A$ if the
magnetic moment $\mu \sim T_{\perp}/B$ is invariant; when the thermal
anisotropy exceeds the firehose threshold, firehose instabilities
will be induced and help suppress the growth of the anisotropy. The
two competing processes will arrive at a balance close to the
threshold, i.e., $A\simeq 1- 2/\beta_{\parallel}$, which gives
$A_1\simeq 0.94$. Introduction of pickup ions in the outer
heliosphere would not significantly change this value. The newly-born
pickup ions gyrate about the interplanetary magnetic field, forming a
ring-beam distribution; this configuration is unstable to the
generation of MHD waves, which scatter the ions quickly toward
isotropy \citep{lee87, bogdan91}. We have also shown that $A_2$ does
not depend on $A_1$ when $2M_{A1}^2 \gg \beta_1$. Substituting the
values of $M_{A1}$, $\beta_1$, $r_s$ and $A_1$ into equations (8) and
(9) gives $A_2 \simeq 1.03$ and $\beta_{\perp 2}\simeq 42.2$. The
threshold of the mirror mode is $1+1/\beta_{\perp 2} \simeq 1.02$,
smaller than the downstream anisotropy. As indicated by Figure~1,
even an $A_2$ slightly larger than 1 can meet the mirror mode onset
at high $\beta_1$, so mirror mode instabilities should be induced in
the heliosheath.

For the upstream temperature, we have used the density-weighted
average of the pickup ion, solar wind proton and electron
temperatures \citep{whang04}. However, the result can be shown to be
self-consistent. Equation~9 is reduced to $\beta_2 \sim
\frac{2M_{A1}^2}{r_s^2}(1-1/r_s)$ in the TS case, from which we have
the downstream temperature $T_2 \sim 2\times 10^6$ K given the
average field magnitude $B_2 \simeq 0.09$ nT and density $n_2 \simeq
0.002$ cm$^{-3}$ from equation~7. Approximating the TS as a
hydrodynamic shock because of the high plasma beta,
$$\frac{T_2}{T_1} = \frac{[2\gamma M^2-(\gamma-1)][(\gamma-1)M^2+2]}
{(\gamma+1)^2M^2},$$ we obtain $T_1 \sim 5\times 10^5$ K, consistent
with the MHD simulation result. Here $\gamma = 5/3$, and $M\simeq 3$,
the shock Mach number estimated from $n_2/n_1 \simeq 3$.

Given the absence of plasma measurements across the TS, it is hard to
estimate the uncertainties brought about by the upstream conditions
and the shock parameters. A revisit to equations~(1)-(6) assuming a
10\% error in the upstream conditions and the shock geometry and
strength gives $A_2 = 1.03\pm 0.42$ and a mirror mode threshold
$1.02\pm 0.01$. Note that the uncertainty of $A_2$ is determined
largely by the shock strength, since $A_2 \sim 1/(r_s-3)$ in the case
of the TS. If we use $r_s = 2.6^{+0.4}_{-0.2}$ inferred from the
particle spectra \citep{stone05} with other parameters fixed, then
the temperature anisotropy $A_2 = 1.03$ - $4.19$; the mirror mode
would be more likely to occur. Interestingly, the shock strength
cannot be smaller than 2.2 at the present conditions; otherwise $A_2$
would be negative. It should be emphasized that turbulence induced by
the mirror mode instability would leave the threshold marginally
satisfied, so we expect $A_2 = 1.02\pm 0.01$ in the evolved state.

The mirror mode instability has maximum growth rate at a wave vector
highly oblique to the background magnetic field. The minimum variance
direction of the measured magnetic fields has an angle of about
75$^{\circ}$ with respect to the background field when magnetic holes
are present \citep{burlaga06b}. Mirror mode waves are non-propagating
and appear to be static structures, consistent with the observed
magnetic fluctuations with a fairly constant field direction
\citep{burlaga06a, burlaga06b}. Mirror mode waves are also
characterized by anti-correlated density and magnetic fluctuations.
The density measurements are not available, so we cannot verify the
mirror mode from density fluctuations. The density perturbation
$\delta n$ can be estimated from \citep{hasegawa69}
$$\frac{\delta n}{n} =
-(\frac{T_{\perp}}{T_{\parallel}}-1)\frac{\delta B}{B},$$ where
$\delta B$ is the perturbation in the field strength. The fluctuation
amplitude $\delta B/B$ is $\sim 0.4$ - 0.9 in the heliosheath, which
results in $\delta n/n \sim 0.01$ - 0.03, too small to be resolved by
future V2 observations.

The non-linear evolution of mirror mode instabilities would also
generate holes in the background field strength \citep{kivelson96}.
The magnetic holes in the heliosheath are of the similar size (in
units of the proton gyroradius) to those in planetary magnetosheaths
that are predominately produced by mirror mode waves
\citep{kaufmann70, bavassano98, burlaga06b}. Magnetic holes in
planetary magnetosheaths have also been explained as slow mode
solitons based on Hall-MHD theory \citep{stasie04}. The plasma beta
in the heliosheath is $\sim$ 40 as estimated above; in this case the
ion Larmor radius is much larger than the ion inertial length, so the
Hall-MHD theory breaks down \citep{pok05}.

Deep in the heliosheath, magnetic field lines may be draped and
compressed against the heliopause if there is no significant
reconnection between the solar wind and interstellar fields.
Analogous to planetary magnetosheaths, the plasma would be squeezed
and flow along the draped field lines, leading to a plasma depletion
layer close to the heliopause. The plasma beta is reduced in this
layer, so the mirror mode instability may be inhibited and ion
cyclotron waves may be generated. Damping of these waves would
suppress the thermal anisotropy and heat the plasma \citep{liu06b}.

\section{Summary}
A simple calculation based on temperature anisotropy instabilities
explains a variety of observations associated with interplanetary and
planetary bow shocks. A shock modifies the velocity distribution of
particles at its surface, producing instabilities downstream of the
shock which give rise to different types of waves. The calculation
also predicts that mirror mode waves form downstream of the TS, which
is consistent with the observed magnetic fluctuations. The present
results provide a substantial basis for shock-induced anisotropies
which act as an energy source for plasma heating in various space and
astrophysical environments.

\acknowledgments The research was supported under NASA contract
959203 from JPL to MIT, NASA grant NAG5-11623, and by the NASA
Planetary Atmospheres and Outer Planets Programs. This work was also
supported by the CAS International Partnership Program for Creative
Research Teams.

\clearpage

\begin{figure}
\plotone{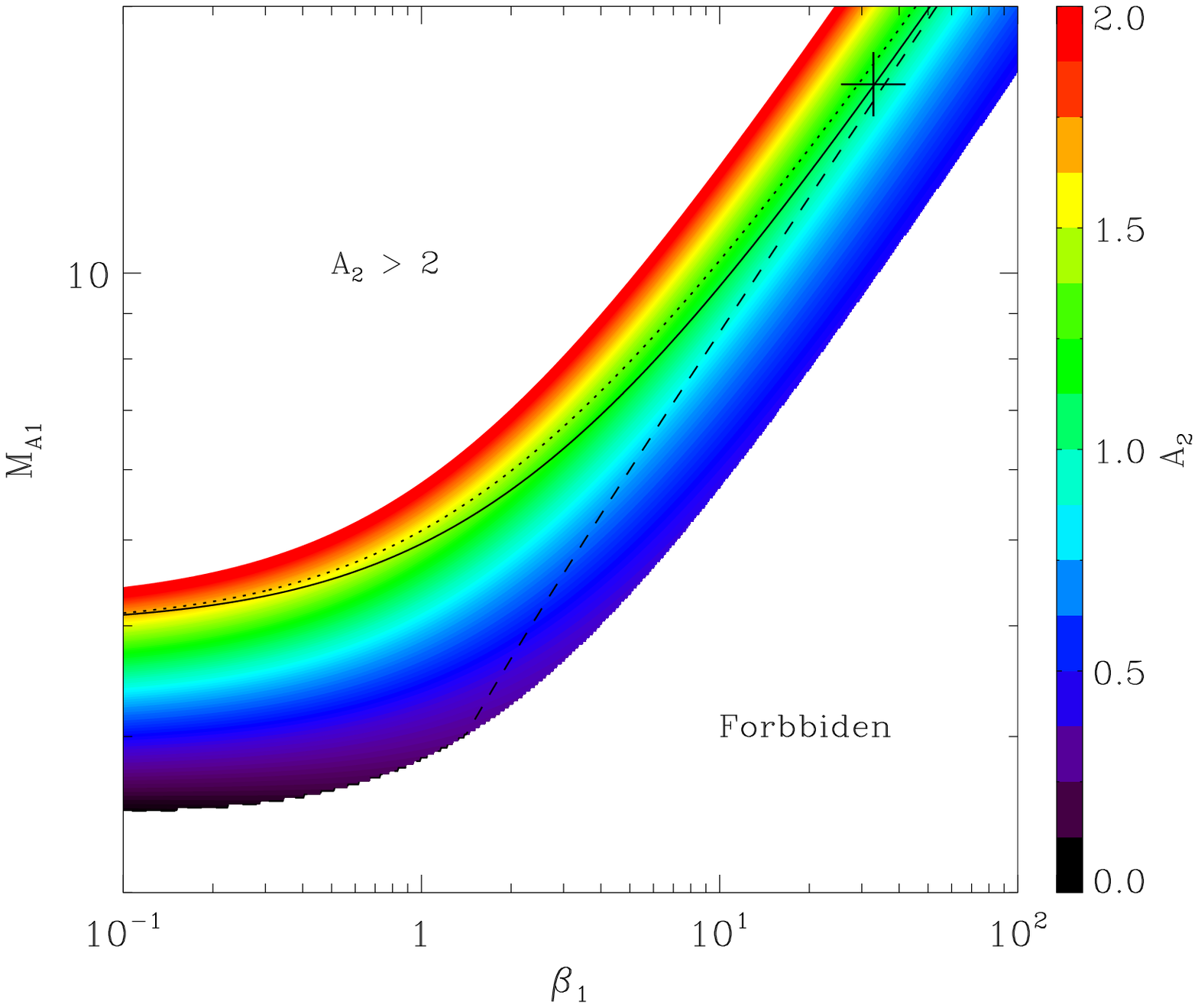}
\caption{Temperature anisotropy downstream of a
perpendicular shock with $r_s=3$ and $A_1=1$ as a function of
$\beta_1$ and $M_{\rm A1}$. The color shading denotes the values of
$A_2$. The lower region is forbidden for a $r_s = 3$ shock since the
entropy does not increase across the shock. Also shown are the
thresholds for the mirror mode (solid line), ion cyclotron (dotted
line) and firehose (dashed line) instabilities. Regions above the
mirror/ion cyclotron threshold are unstable to the mirror/ion
cyclotron mode and regions below the firehose onset are unstable to
the firehose instability. The plus sign marks the TS location.}
\end{figure}

\clearpage

\begin{figure}
\plotone{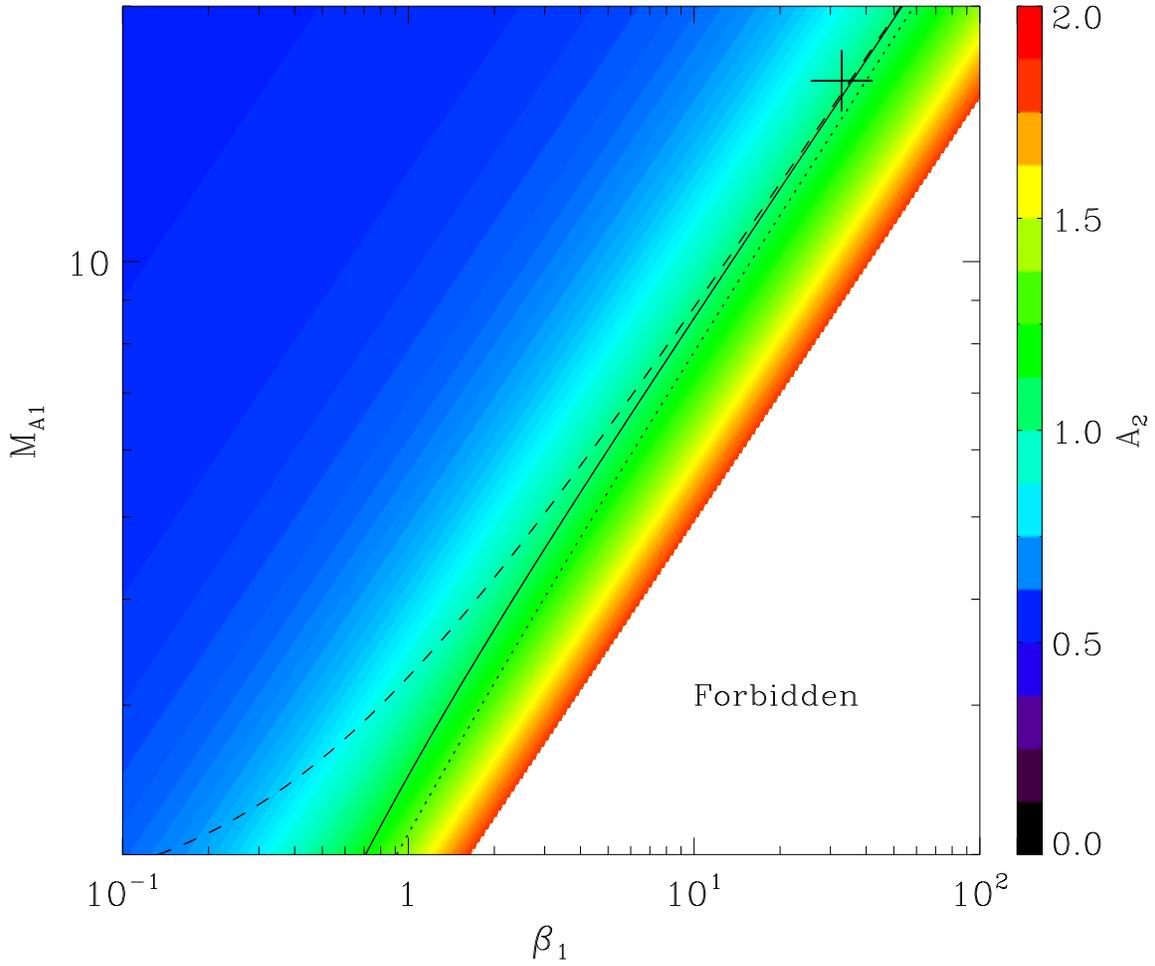}
\caption{Temperature anisotropy downstream of a
parallel shock with $r_s=3$ and $A_1=1$ as a function of $\beta_1$
and $M_{\rm A1}$. Same format as Figure~1. Regions below the
mirror/ion cyclotron threshold are unstable to the mirror/ion
cyclotron mode and regions above the firehose onset are unstable to
the firehose instability.}
\end{figure}

\end{document}